\documentclass[a4paper]{article}
\usepackage{multirow}
\usepackage{url}
\usepackage{cite}
\usepackage{INTERSPEECH2020}
\usepackage{setspace}
\usepackage{colortbl}

\title{Efficient Integration of Multi-channel Information for Speaker-independent Speech Separation}
\name{Yuichiro Koyama$^{12}$, Oluwafemi Azeez$^{13}$, Bhiksha Raj$^1$}
\address{
  $^1$Carnegie Mellon University, Pittsburgh, PA, USA\\
  $^2$Sony Corporation, Tokyo, Japan\\
  $^3$InstaDeep, Lagos, Nigeria}
\email{Yuichiro.Koyama@sony.com, f.azeez@instadeep.com, bhiksha@cs.cmu.edu}

\begin{document}
\setlength{\abovedisplayskip}{5pt} 
\setlength{\belowdisplayskip}{5pt} 

\maketitle
\begin{abstract}
Although deep-learning-based methods have markedly improved the performance of speech separation over the past few years, 
it remains an open question how to integrate multi-channel signals for speech separation.
We propose two methods, namely, early-fusion and late-fusion methods, to integrate multi-channel information based on the time-domain audio separation network, which has been proven effective in single-channel speech separation.
We also propose channel-sequential-transfer learning, which is a transfer learning framework that applies the parameters trained for a lower-channel network as the initial values of a higher-channel network.
For fair comparison, we evaluated our proposed methods using a spatialized version of the wsj0-2mix dataset, which is open-sourced.
It was found that our proposed methods can outperform multi-channel deep clustering and improve the performance proportionally to the number of microphones.
It was also proven that the performance of the late-fusion method is consistently higher than that of the single-channel method regardless of the angle difference between speakers.
\end{abstract}

\section{Introduction}
Speech separation is necessary to improve the performance of other speech-related technologies (e.g., speech recognition and speaker diarization) when there are multiple speakers talking simultaneously~\cite{watanabe2020chime}. This so-called cocktail-party problem~\cite{cherry1953some,bregman1994auditory}
is still a challenging problem even though it has been intensively researched for a long time~\cite{ellis1996prediction,choi2005blind,hiroe2006solution,schmidt2006single,smaragdis2007conv,rouat2008computational,le2015sparse,wood2017blind}.

Deep-learning-based methods have markedly improved the performance of speech separation in recent years. In the domain of single-channel speech separation, methods that estimate time-frequency masks, such as deep clustering (DC)~\cite{hershey2016deep,isik2016single}, permutation-invariant training (PIT)~\cite{yu2017permutation, kolbaek2017multitalker}, 
computational auditory scene analysis (CASA)-based approaches~\cite{liu2018casa},
and the deep attractor network~\cite{luo2018speaker}, have achieved a high level of success. 
However, these approaches have limited performance since they only determine the spectral amplitudes of the target speech without accounting for the phase, which causes the inconsistency of the short-time Fourier transform (STFT)~\cite{griffin1983signal,wisdom2019differentiable}. 
On the other hand, the time-domain audio separation network (TasNet)~\cite{luo2018tasnet}, which has a trainable encoder-decoder architecture and does not depend on the STFT, avoids this inconsistency. In particular, Conv-TasNet~\cite{luo2019conv}, a fully convolutional version of TasNet with a temporal convolutional network (TCN)~\cite{lea2016temporal,lea2017temporal,bai2018empirical}, has surpassed the performance of the ideal binary mask (IBM), which is generally considered as the upper limit of the methods that depend on time-frequency masks.

Approaches that utilize multiple microphones such as a microphone array --  so-called multi-channel speech separation approaches -- are generally more effective than those that use a single microphone, since the recorded signals also carry distinguishing spatial information about the sources. While deep learning can also be applied to multi-channel speech separation, proposed solutions still have problems with effectively integrating the multi-channel information, especially with more than two microphones. For example, the performance of multi-channel deep clustering (MCDC)~\cite{wang2018multi}, an extended version of DC with spatial information added to the input of DC, is higher than that of DC for a single-channel. MCDC, however, only considers signals from {\em pairs} of microphones at a time, regardless of the number of available microphones. While predictions obtained with different pairs of microphones may be merged, the computational complexity increases proportionally to the number of pairs of microphones. A similar approach in terms of feeding signals from a pair of microphones into the network has also been proposed in \cite{wang2018integrating} and has the same problem as MCDC. Other multi-channel methods utilize the output of conventional integration methods (e.g., a beamformer or multi-channel Wiener filter) as the input to a network~\cite{chen2018efficient,togami2019spatial}. However, these methods are arguably not {\em true} multi-channel deep learning approaches, since the multi-channel signals are already combined into a single signal through a separately optimized process prior to being input to the network; the network itself effectively merely acts as a learned post-filter. In \cite{chen2019multi,gu2019neural}, every inter-channel phase difference (IPD) calculated from each pair of microphones is utilized as the input of the network. However, a network that uses only the IPD for the input has inferior performance to even the single-channel model when the angle difference between speakers is small and additional input features such as the output of the beamformer are required to solve this problem~\cite{gu2019neural}. Furthermore, as for the time-domain method, the performance of the multi-channel method is often lower than that of the single-channel method as shown in \cite{luo2019fasnet}. 
In addition, including recent end-to-end approaches~\cite{gu2019end,e2e2020Zhang}, these approaches are evaluated using different datasets, so comparing their performance is difficult. In summary, it remains an open question how to integrate multi-channel information for speech separation within a deep learning framework.

In this paper, we propose two extensions of the single-channel Conv-TasNet algorithm to perform {\em true} multi-channel speech separation. The first method, which we call {\em early-fusion}, integrates signals prior to being processed by the separation block, while the second, which we term {\em late-fusion}, extracts component signals from the individual channels before combining them. The multi-channel aspect of the model is achieved through the learning procedure, which jointly optimizes {\em all} components of the processing for signal-separation performance. Finally, we also propose a transfer-learning framework: {\em channel-sequential-transfer learning (CSTL)}, for optimal initialization of model parameters for both approaches, prior to optimization.
The inference-time computational complexity of the early-fusion method is almost independent of the number of microphones because the calculation of the separation block is performed only once. Computation in late-fusion scales linearly with the number of microphones.

Experimental evaluations on the dataset utilized in \cite{wang2018multi}, which is open-sourced, lead to the following conclusions:
First, the performance of the early-fusion method is extremely high in an anechoic environment, even higher than that of the IBM-based minimum variance distortionless response (MVDR) beamformer.
Second, both early-fusion and late-fusion methods, combined with CSTL, outperform MCDC under reverberation conditions. Moreover, their performance improves proportionally to the number of microphones.
Third, the performance of the late-fusion method is consistently higher than that of the single-channel method regardless of the angle difference between speakers, while the performance of the early-fusion method can be lower than that of the single-channel method when the difference is small.

\section{Related work}
\subsection{Problem definition}
Multi-channel speech separation operates on signals captured by an array of $M$ microphones and tries to separate $K$ mixed speech signals. 
The discrete-time signal captured by the $m$th microphone in the array can be written as 
\begin{equation}
\begin{split}
x_m(n) &= \sum^K_{k=1}a_{m,k}(n)*s_{0k}(n)+v_m(n) \\
&=\sum^K_{k=1}s_{m,k}(n)+v_m(n),   m=1,2,\dots,M,
\end{split}
\label{eq:observation}
\end{equation}
where $*$ is the convolution operator, 
$a_{m,k}$ the channel impulse
response between the $k$th speech and the $m$th microphone,
$s_{0k}$ the $k$th speech signal without reverberation,
$s_{m,k}$ the reverberant speech component, 
and $v_m$ the noise at the $m$th microphone.

In this paper, we assume that the number of sources, $K$, is given, and that the goal of multi-channel speech separation is defined as estimating $s_{r,k}$, where $r$ is the desired channel.
Hereafter, $r$ is replaced with $1$ without loss of generality and $v_m(n)$ is omitted, assuming a noise-free situation. 

\subsection{Conv-TasNet}
We will now review Conv-TasNet~\cite{luo2019conv}.
Assuming that the number of microphones available is only one ($m$ is omitted) in Eq.~\eqref{eq:observation},
the discrete-time signal captured by the microphone $x(n)$ can be divided into overlapping segments of length $L$, represented by $\mathbf{x}_t\in\mathbb{R}^{L}$,
where $t = 1,\dots,T$ represents the segment index and $T$ represents the total number of segments.
A matrix $\mathbf{X}\in \mathbb{R}^{L\times T}$ can then be formed by concatenating $\mathbf{x}_t$ for all segments $t$.
$\mathbf{X}$ is transformed into $N$-dimensional representations $\mathbf{W}\in\mathbb{R}^{N\times T}$ for all segments by multiplying by a trainable linear encoder $\mathbf{U}\in \mathbb{R}^{N\times L}$ as follows:
\begin{equation}
\mathbf{W} = \mathbf{U}\mathbf{X}.
\label{eq:encoder}
\end{equation}
$\mathbf{W}$ is first fed into a bottleneck layer (BNL), which performs global layer normalization (gLN) and a $1\times1$ convolution, 
\begin{equation}
\mathbf{B} = \text{BNL}(\mathbf{W}),
\label{eq:bnl}
\end{equation}
where $\mathbf{B}\in\mathbb{R}^{B\times T}$ is the bottleneck feature and $B$ is the number of channels of the bottleneck layer. Then, $\mathbf{B}$ is fed into a TCN block~\cite{lea2016temporal,lea2017temporal,bai2018empirical} as
\begin{equation}
\mathbf{Y} = \text{TCN}(\mathbf{B}),
\label{eq:tcn}
\end{equation}
where $\mathbf{Y}\in \mathbb{R}^{B\times T}$ is the feature embedding calculated by accumulating the outputs of all convolutional blocks based on depthwise separable convolutions~\cite{howard2017mobilenets} in the TCN. The masks for $K$ speech signals $\mathbf{M}_k\in \mathbb{R}^{N\times T} (k=1,\dots,K)$ are calculated by mask estimation (ME) blocks, which are composed of the parametric rectified linear unit (PReLU), a 1$\times$1 convolution, and a sigmoid function, and then multiplied by $\mathbf{W}$ as
\begin{equation}
\mathbf{M}_k = \text{ME}_k(\mathbf{Y}),
\label{eq:me}
\end{equation}
\begin{equation}
\mathbf{Z}_k = \mathbf{M}_k\odot\mathbf{W},
\label{eq:masking}
\end{equation}
where $\mathbf{Z}_k\in \mathbb{R}^{N\times T}$ is the $N$-dimensional representation of each speech signal
and $\odot$ is the Hadamard product.
$\mathbf{Z}_k$ is multiplied by a trainable linear decoder $\mathbf{V}\in \mathbb{R}^{L\times N}$,
\begin{equation}
\hat{\mathbf{S}}_k = \mathbf{V}\mathbf{Z}_k,
\label{eq:decorder}
\end{equation}
where $\hat{\mathbf{S}}_k\in \mathbb{R}^{L\times T}$
is each estimated speech signal for all segments.
Each estimated speech signal $\hat{s}_k(n)$ is finally reconstructed by overlapping and adding the $T$ columns in $\hat{\mathbf{S}}_k$.

The parameters of the encoder, BNL, TCN,
ME, and decoder are learned by minimizing the scale-invariant source-to-noise ratio (SI-SNR) loss $L_{\text{SI-SNR}}$, which is defined as
\begin{equation}
L_{\text{SI-SNR}} = -\frac{1}{K}\sum_{k=1}^K10\log_{10}(\|\alpha\mathbf{s}_k\|^2/\|\alpha\mathbf{s}_k-\hat{\mathbf{s}}_k\|^2),
\label{eq:sisnrloss}
\end{equation}
where $\mathbf{s}_k$ and $\hat{\mathbf{s}}_k$
are vector representations of $s_k(n)$ and $\hat{s}_k(n)$, respectively, and $\alpha = \langle \mathbf{s}_k,\hat{\mathbf{s}}_k \rangle/\|\mathbf{s}_k\|^2$.
To solve the permutation problem, utterance-level PIT is applied~\cite{kolbaek2017multitalker}.

\section{Proposed method}
\begin{figure*}[htb]
  \centering
  \includegraphics[width=15.0cm]{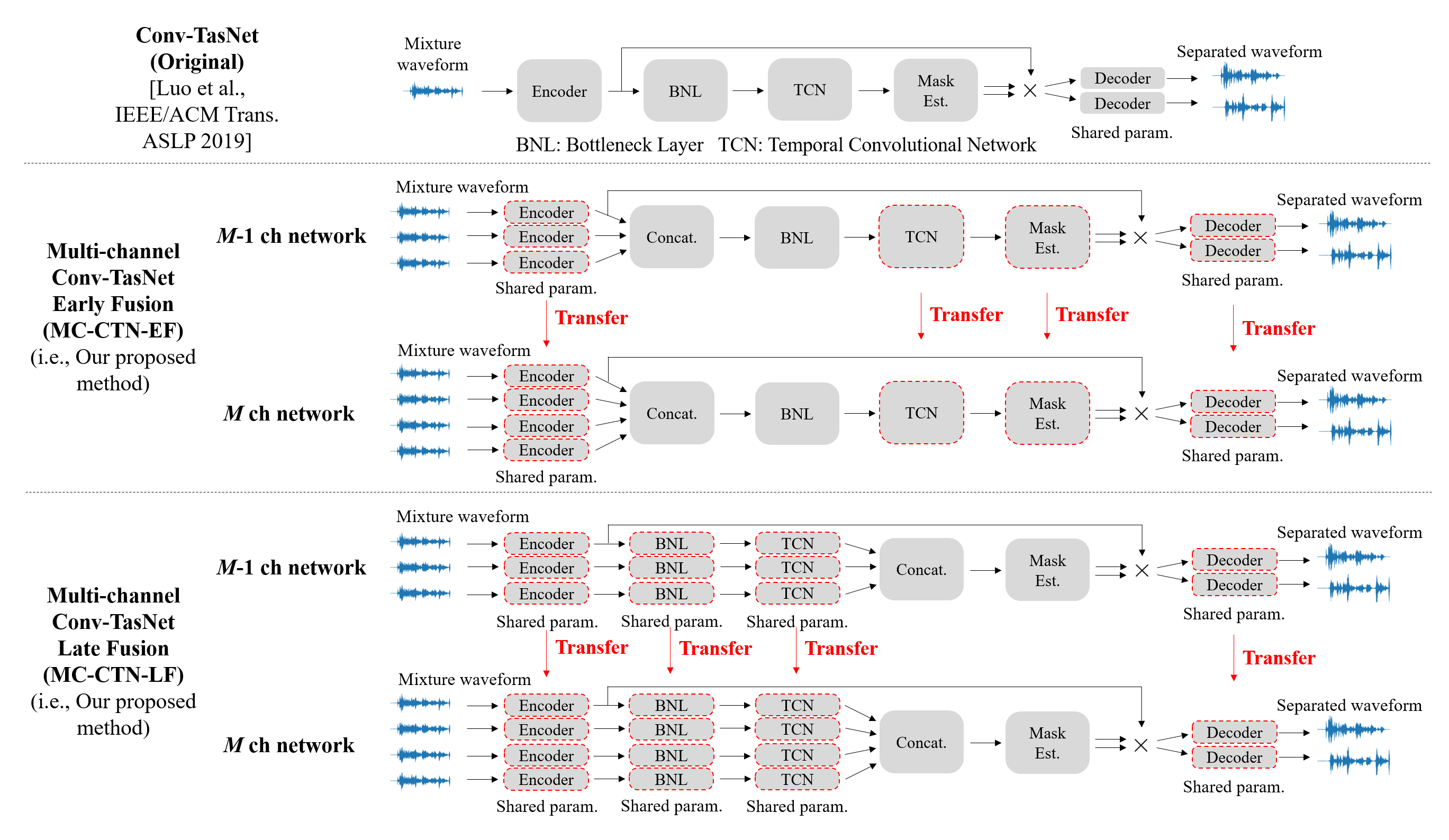}
  \caption{Our proposed methods. In the early-fusion method, each block except for the BNL is independent of the number of microphones. In the late-fusion method, each block except for the ME is independent of the number of microphones. The parameters obtained when training an $M-1$ ch network are utilized as the initial values of the $M$ ch network.
}
\label{fig:proposed_method}
\vspace{-3mm}
\end{figure*}
We propose to extend Conv-TasNet to a multi-channel version in two ways (i.e., early- and late-fusion methods). 
Figure~\ref{fig:proposed_method} illustrates our proposed methods.
In either case, Eq.~\eqref{eq:encoder} is extended to the following multi-channel version by adding a subscript $m$, except to the encoder $\mathbf{U}$,
\begin{equation}
\mathbf{W}_m = \mathbf{U}\mathbf{X}_m.
\label{eq:encoder_mc}
\end{equation}

\subsection{Early-fusion method}
In the early-fusion method, 
multi-channel information is integrated before being fed into the BNL.
Specifically, $\mathbf{W}_0\in\mathbb{R}^{MN\times T}$ is computed by concatenating $\mathbf{W}_m$ in terms of all microphones and fed into $\text{BNL}_{\text{EF}}$, which is the BNL extended such that the $MN\times T$ matrix can be dealt with,
\begin{equation}
\mathbf{B} = \text{BNL}_{\text{EF}}(\mathbf{W_0}).
\label{eq:tcn_ef}
\end{equation}
Then, each estimated speech signal, $\hat{s}_{1,k}(n)$, is obtained in a similar manner from Eqs.~\eqref{eq:tcn} to \eqref{eq:decorder}.
Note that each block except for the BNL is independent of the number of microphones in the early-fusion method.
We will refer to this approach as multi-channel Conv-TasNet early fusion (MC-CTN-EF).

\subsection{Late-fusion method}
In the late-fusion method, 
multi-channel information is integrated after being fed into the TCN block.
Specifically, 
BNL and TCN blocks, whose parameters are shared with all microphones, are first processed,
\begin{equation}
\mathbf{Y}_m = \text{TCN}(\text{BNL}(\mathbf{W}_m)).
\label{eq:tcn_lf}
\end{equation}
Then, $\mathbf{Y}_0\in\mathbb{R}^{MB\times T}$ is computed by concatenating $\mathbf{Y}_m$ for all microphones and fed into $\text{ME}_{\text{LF}}$ block, which is the ME block extended such that the $MB\times T$ matrix can be dealt with,
\begin{equation}
\mathbf{M}_k = \text{ME}_{\text{LF},k}(\mathbf{Y}_0).
\label{eq:me_lf}
\end{equation}
Each estimated speech signal $\hat{s}_{1,k}(n)$ is obtained in a similar manner from Eqs.~\eqref{eq:masking} and \eqref{eq:decorder}.
Note that each block except for ME is independent of the number of microphones in the late-fusion method.
We will refer to this approach as multi-channel Conv-TasNet late fusion (MC-CTN-LF).

\subsection{Channel-sequential-transfer learning}
Since most of the blocks in our proposed methods are independent of the number of microphones as mentioned above, the parameters obtained when training the network for a lower number of microphones are utilized as the initial values. 
Specifically, we propose channel-sequential-transfer learning (CSTL), which is a transfer learning that applies the parameters trained for an $M-1$ channel network as the initial values for an $M$ channel network as shown in Figure~\ref{fig:proposed_method}.
These approaches do not significantly increase the number of trainable parameters even if the number of available microphones increases because most of the blocks are independent of the number of microphones.
In particular, in the early fusion method, the computational complexity also does not significantly increase because the calculation of the TCN block is performed only once.

\section{Experiment}
\begin{table*}
\begin{center}
    \caption{Evaluation results obtained using the spatialized version of the wsj0-2mix dataset. SDR and SI-SNRi are shown in dB scale. By CSTL, the performance characteristics of both MC-CTN-EF and MC-CTN-LF were improved proportionally to the number of microphones, and both outperformed MCDC and single-channel Conv-TasNet.}
  \label{tab:result}
 \scalebox{0.78}{
 
    \centering
    \begin{tabular}{l|c|ccc|ccc|ccc|ccc}
    \toprule
        \multicolumn{2}{c|}{}  & \multicolumn{3}{c|}{Anechoic 2ch} & \multicolumn{3}{c|}{Reverberation 2ch} &
\multicolumn{3}{c|}{Reverberation 3ch} &
        \multicolumn{3}{c}{Reverberation 4ch} \\ 
        Method & Pre-train & SDR & SI-SNRi & PESQ  & SDR & SI-SNRi & PESQ  & SDR & SI-SNRi & PESQ & SDR & SI-SNRi & PESQ \\
        \midrule
        MCDC~\cite{wang2018multi} & - & 12.9  & - & - & 8.9  & - & - & 9.3  & - & - & 9.4  & - & - \\ 
        MCDC (our impl.) & - & 12.54  & 12.04  & 2.73  & 9.44  & 8.87  & 2.39  & 9.77  & 9.22  & 2.42  & 9.87  & 9.32  & 2.43  \\ 
        \midrule
        Conv-TasNet (Ch.1) & - & 14.01  & 13.60  & 3.01  & 9.50  & 8.92  & 2.80  & - & - & - & - & - & - \\ 
        \midrule
        MC-CTN-EF & - & 25.38  & 24.97  & \bf{3.96}  & 8.65  & 8.03  & 2.61  & 9.27 & 8.64 & 2.66 & 9.54  & 8.89  & 2.68  \\ 
         & 1ch & \bf{25.45} & \bf{25.03} & \bf{3.96}  & \bf{10.07}  & \bf{9.49}  & 2.85  & 9.74  & 9.17  & 2.83  & 9.65  & 9.06  & 2.80  \\ 
         & 2ch & - & - & - & - & - & - & \bf{10.44}  & \bf{9.87}  & 2.86  & 10.33  & 9.74  & 2.86  \\ 
         & 3ch & - & - & - & - & - & - & - & - & - & \bf{10.60}  & \bf{10.03}  & 2.88  \\ 
         \midrule
        MC-CTN-LF & - & 15.21  & 14.80  & 3.11  & 8.10  & 7.49  & 2.62  & 8.86 & 8.28 & 2.69 & 6.65 & 6.05 & 2.45 \\ 
         & 1ch & 14.21  & 13.80  & 3.03  & 10.03  & 9.45  & \bf{2.86}  & 10.04  & 9.46  & 2.86  & 9.86  & 9.31  & 2.84  \\ 
         & 2ch & - & - & - & - & - & - & 10.21  & 9.65  & \bf{2.88}  & 10.04  & 9.48  & 2.86  \\
         & 3ch & - & - & - & - & - & - & - & - & - & 10.32  & 9.77  & \bf{2.90}  \\ 
         \midrule
        IBM MVDR & - & 21.59  & 19.78  & 3.58  & 5.55  & 3.86  & 2.49  & 8.54 & 5.79 & 2.70 & 10.05 & 6.62 & 2.84 \\ 
        IBM (Ch.1) & - & 13.52  & 13.01  & 3.27  & 12.73  & 12.22  & 3.24  & - & - & - & - & - & - \\ 
        \bottomrule
    \end{tabular}

}
\end{center}
\vspace{-3mm}
\end{table*}

\begin{figure*}[htb]
  \centering
  \includegraphics[width=17.0cm]{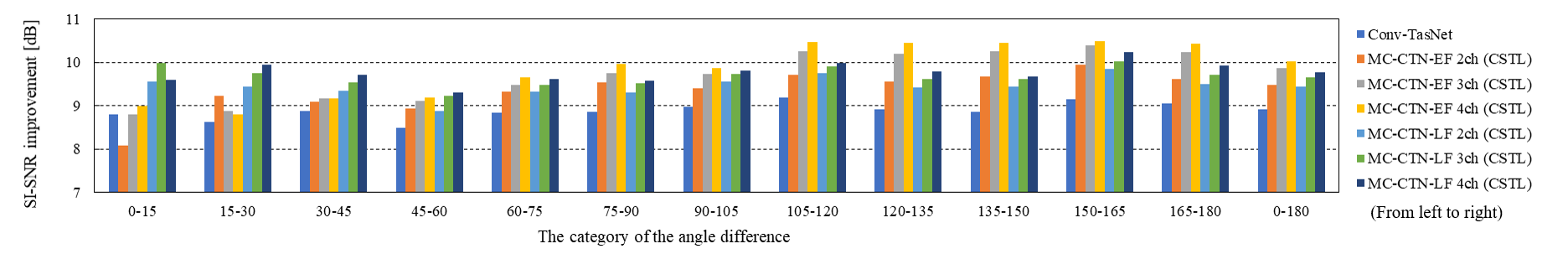}
  \caption{Evaluation results obtained in the sets divided into 12 categories with the angle difference of speakers categorized into ranges of 15 degrees. The performance of MC-CTN-LF was consistently higher than that of single-channel Conv-TasNet regardless of the angle difference of the speakers and the number of channels, while the performance of MC-CTN-EF for 2ch was lower than that of the single-channel Conv-TasNet when the angle difference was small (0--15 degrees).}
\label{fig:result_ad}
\vspace{-5mm}
\end{figure*}

\subsection{Experimental design}
For fair comparison, we evaluated our proposed methods using a spatialized version of the wsj0-2mix dataset, which is open-sourced\footnote{\url{
https://www.merl.com/demos/deep-clustering
}}.
We used the same configuration as in \cite{wang2018multi}. Specifically, the training and validation sets were synthesized by mixing utterances randomly selected from different speakers in the WSJ0 training set such that the signal-to-noise ratio (SNR) was distributed between -5~dB and +5~dB. 
The test set was similarly synthesized using the speakers in the development and evaluation sets of the WSJ0 corpus, which means that the task we solved was speaker-independent speech separation since the test speakers were unseen in the training and validation sets. 
There were 20,000, 5,000, and 3,000 utterances in the training, validation, and test sets, respectively. The speaker locations and microphone geometry were randomly sampled such that the minimum distance between microphones was 5~cm, maximum distance was 25~cm, the minimum distance between speakers was 1~m, and minimum distance between a speaker and the center of the microphone array was 50~cm. Note that the microphone geometry was different from sample to sample. 
The anechoic and reverberation sets were separately generated, and the reverberation time T60 was randomly sampled from 0.2~s to 0.6~s in the reverberation set.
The sampling rate was 8~kHz.
We focused on the situation where the number of speakers was two ($K=2$) in this experiment.

All networks in this experiment were trained on 4-second-long segments using the Adam~\cite{kingma2014adam} optimizer.
The learning rate was set to 0.001.
The training iteration was halted when the average of $L_{\text{SI-SNR}}$ in terms of the validation set did not improve in six consecutive epochs.
The hyperparameters of the network were set as 
$L=16$, $N=512$, and $B=128$, which are based on the best configuration of the original Conv-TasNet~\cite{luo2019conv}.

We utilized SDR calculated by the BSS Eval toolbox~\cite{vincent2006performance}, SI-SNR improvement (SI-SNRi), and PESQ~\cite{rix2001perceptual} to evaluate our methods.
MCDC and Conv-TasNet (for a single channel) were also implemented and evaluated for comparison.
The single-channel signal enhanced by the IBM and 
minimum variance distortionless response (MVDR) beamformer~\cite{souden2009optimal} using the IBM (i.e., oracle-mask MVDR)
were also evaluated.

\subsection{Results and discussion}
The evaluation results are shown in Table~\ref{tab:result}.
The results for MCDC in the first row were copied from the original paper~\cite{wang2018multi}, and those in the second row were obtained by ourselves, which were similar to the original results.
First, Conv-TasNet outperformed MCDC in the 2ch dataset even though it utilized only a single-channel input. It also surpassed IBM, which is consistent with the result of \cite{luo2019conv}. 
For the anechoic set, both MC-CTN-EF and MC-CTN-LF trained from scratch outperformed MCDC and single-channel Conv-TasNet. 
In particular, the performance of MC-CTN-EF was extremely high and surpassed MVDR using the IBM, which is a similar tendency with the result shown in \cite{e2e2020Zhang}.
For the reverberation set, neither MC-CTN-EF nor MC-CTN-LF trained from scratch surpassed single-channel Conv-TasNet, which is also a similar tendency with the result shown in \cite{luo2019fasnet}.
By CSTL, however, the performance characteristics of both methods were improved proportionally to the number of microphones, and they outperformed MCDC and single-channel Conv-TasNet.
Interestingly, although MC-CTN-EF had better performance than MC-CTN-LF in terms of SDR and SI-SNRi, MC-CTN-LF consistently had  better performance than MC-CTN-EF in terms of PESQ.
The performance of the MVDR using IBM was not so high in the reverberation set, which is a similar tendency to the result of \cite{wang2018multi}. 
Note that both methods can work on any microphone array since the microphone geometry was different from sample to sample in both the training set and the test set as mentioned above.
Although the number of trainable parameters of MCDC was 33.78 million, those of 2ch MC-CTN-EF, 3ch MC-CTN-EF, and 4ch MC-CTN-EF were 5.05, 5.12 5.18 million respectively and those of 2ch MC-CTN-LF, 3ch MC-CTN-LF, and 4ch MC-CTN-LF were 5.18, 5.31, 5.44 respectively, indicating that MC-CTN-EF and MC-CTN-LF are efficient in integrating multi-channel signals.

Figure~\ref{fig:result_ad} shows the average of SI-SNRi in the sets divided into 12 categories with the angle difference of speakers categorized into ranges of 15 degrees. 
It was found that the performance of MC-CTN-LF was consistently higher than that of single-channel Conv-TasNet regardless of the angle difference of the speakers and the number of channels, while the performance of MC-CTN-EF for 2ch was lower than those of the single-channel Conv-TasNet when the angle difference was small (0--15 degrees).

\section{Conclusions}
In this paper, we proposed two methods of integrating multi-channel information for speech separation, namely MC-CTN-EF and MC-CTN-LF. 
MC-CTN-EF do not significantly increase the computational complexity even if the number of available microphones increases since the separation block is processed only once.
We also proposed a transfer learning called CSTL, which improved the performance under a reverberation condition and outperformed MCDC.
In addition, the performance of MC-CTN-LF was consistently higher than that of single-channel Conv-TasNet regardless of the angle difference of the speakers.  
These findings are helpful in developing real applications since observed signals often include reverberation, and speakers can talk to a device from any directions under real-life conditions. 
We will focus on a dataset that includes noise signals such as that in \cite{WHAMR} and the situation where more than two speakers are talking simultaneously in future work.

\bibliographystyle{IEEEtran}

\bibliography{mybib}

\end{document}